# Characterization of electron detectors by time-of-flight in neutron $\beta$-decay experiments

Dirk Dubbers

*Physikalisches Institut der Universität, Im Neuenheimer Feld 226, 69120 Heidelberg, Germany*

ABSTRACT

Progress in neutron decay experiments requires better methods for the characterization of electron detectors. I show that for such $\beta$-decay studies, electron time-of-flight can be used for in-situ calibration of electron detectors. Energy resolution down to a few keV can be reached for the lower part of the electron spectrum in neutron decay, where conventional calibration methods come to their limit. Novel time-of-flight methods can also be used to perform a complete experiment on electron backscattering from their detectors.

*Keywords:*
Beta decay
Electron detectors
Neutrons
Charged-particle spectroscopy
Nuclear tests of fundamental interactions and symmetries

# 1. Introduction

Progress on the precision frontier of low-energy particle physics relies to a large extent on progress in instrumentation. Experiments on nuclear and neutron *β*-decay, searching for new physics beyond the standard model, nowadays aim at $10^{-4}$ precision in the accessible decay parameters, see the recent reviews [1- 6]. In neutron decay, such parameters are, besides the neutron lifetime, the electron-neutrino correlation, the electron-, proton-, neutrino-asymmetries, and others as listed in [7]. This ambitious aim requires, among other challenges, the development of improved methods of detector characterization.

Main issues in the characterization of electron detectors are (A) energy calibration and (B) backscatter suppression. With new neutron decay spectrometers becoming operational (for a list of neutron decay instruments, see [4]), contemporary calibration methods (A) will soon reach their limits. Problems of conventional energy calibration with conversion electrons, in particular in the lower part of the energy spectrum, include non-linear detector response, self-absorption in the source, and detector surface effects.

Electron backscattering from the detector (B) is also a notorious source of error. Numerous experiments have been done on this topic, see [8,] and references therein. Backscattering coefficients should ideally be known in dependence of scattering angles and energies, both of the in and outgoing electrons. Existing such data sets are far from being complete, while simulated data, which have seen much progress lately, do not cover glancing incidence of the electrons [9].

The topic of the present article is the use of electron time-of-flight (ToF) measurements for detector characterisation in nuclear *β*-decay studies. ToF methods require flight times that are long compared to the intrinsic time resolution of the detectors. This requirement is not met for typical *β*–decay spectrometers, with electron flight times of a few nanoseconds for flight paths of the order of one meter. Therefore, in the past, ToF methods have played no role in *β*-decay studies. Furthermore, in many *β*-decay experiments, magnetic fields are used to guide the decay electrons to their energy sensitive detectors. Due to electron gyration about the guiding field, flight times then depend not only on energy, but also on the angle of electron emission, and the two quantities cannot be disentangled. Magnetic guiding fields are also used in almost all neutron decay spectrometers, for a list see [10].

In the past, electron ToF spectroscopy was possible only in the eV to keV energy range, and has become a standard method in photoelectron spectroscopy [11]. In a more special application, electron flight times are employed in the characterization [12] of the 18 keV tritium *β*-retardation spectrometer KATRIN, searching for a non-zero neutrino mass [13]. At the other extreme of the energy spectrum, in high energy physics, flight times of relativistic electrons with velocity $v \approx c$ are independent of energy, and therefore ToF cannot be used for energy measurement, but only for discrimination of electrons from non-relativistic heavy baryons.

In neutron decay experiments, in situ calibration by electron ToF (A), would be a useful complement to calibration with conversion electrons, if its dependence an electron emission angle can be eliminated. For calibration by ToF, an uncalibrated source of electrons is sufficient. Furthermore, ToF offers high resolution at low electron energies, where flight times are long, in contrast to calibration with conversion electrons, where resolution decreases with decreasing energy.

A complete experiment on electron backscattering (B) can be be done by ToF in a uniform magnetic guiding field. ToF and energy measurements on the incoming and backscattered electron would give backscattering coefficients simultaneously for all angles and energies of the incident and outgoing electrons, all in one single data set.

In the following Section 2, electron ToF spectra under magnetic guidance will be calculated, for simplicity starting with the case of a uniform guiding field. Section 3 first lists existing methods of energy calibration of electron detectors (A). The calculations of Section 2 then are extended to ToF spectra in strongly non-uniform guiding fields, as needed for energy calibration by ToF. Section 4 discusses the use of ToF for the measurement of backscattering coefficients (B). In the last Section 5, I discuss some technical issues, including the generation of a start signal, required for any ToF measurement.

## 2. Electron ToF spectra for a uniform magnetic guiding field

To investigate our topics A and B, we need to calculate the shapes of electron ToF spectra, following transport through non-uniform magnetic guiding fields. One should think that this is a straight-forward and well known exercise. It turns out, however, that this calculation is rather involved, which may be the reason why ToF spectra found in the literature are all calculated with Monte Carlo (MC) ray tracing techniques, even for simple field configurations, as described, for instance, in [14,15].

Monte Carlo simulations are indispensible for the development of intricate instruments, but there are good reasons to accompany MC by analytical studies. First, it is helpful to thoroughly understand a problem before engaging into MC simulations. Preceding papers on the *spatial* distribution of electrons after magnetic transport [16,17] (in contrast to their distribution in *time*, topic of the present paper) had shown that main features of the electron point spread function (namely, the occurrence of an infinite number of unexpected singularities) had remained undetected during decades of MC studies.

A second reason for analytical studies is that ray tracing calculations, even with advanced programs and computers, still are rather slow. For the KATRIN instrument, for example, ray tracing of electron trajectories through the entire spectrometer takes six minutes of CPU time per emitted electron [18] (N.B., this is not the time required to calculate one point of the spectrum, but to add one count to it). This state of affairs requires much patience if spectral and spatial distributions must be studied for a large variety of experimental conditions.

I start with the calculation of ToF spectra with a *uniform* guiding field, which already contains the essence of our method of calculation. The flight time $t$ of an electron in a uniform magnetic field depends on both, the electron's kinetic energy $E$, and its polar emission angle $\theta$ with respect to the field axis $z$,

$$t = z_0 / (v \cos\theta), \tag{1}$$

with the linear distance $z_0$ between detector and source. The velocity $v$ of the electron, divided by the speed of light $c$, is related to energy $E$ as

$$\beta = v/c = cp/W, \tag{2}$$

with electron momentum $p = c\sqrt{E(E + 2m_0 c^2)}$ and electron rest mass $m_0$, which is derived from total energy $W = E + m_0 c^2 = \gamma m_0 c^2$ with the Lorentz factor $\gamma = 1/\sqrt{1-\beta^2}$.

The probability $dP$ that an electron arrives at the detector within an infinitesimal time interval from $t$ to $t + dt$ depends on the corresponding increments of both energy and solid angle. For the case of spatial rotational symmetry, the latter increment is $d\Omega = 2\pi \, d\cos\theta$. Hence,

$$\frac{dP}{dt} = \frac{dP}{dE}\frac{dE}{dt} + \frac{dP}{d\cos\theta}\frac{d\cos\theta}{dt}, \tag{3}$$

Therein, $dP/dE$ is the energy spectrum of the electrons, and $dP/d\cos\theta$ their angular distribution.

The term $dE/dt$ in Eq. (3) is obtained from the derivative of $t$ with respect to $E$, using Eq. (1), whose inverse is

$$\frac{dE}{dt} = -\frac{c}{z_0}\frac{(cp)^3}{(m_0 c^2)^2}\cos\theta. \tag{4}$$

Similarly, the derivative of $t$ with respect to $\cos\theta$ gives

$$\frac{d\cos\theta}{dt} = -\frac{c}{z_0}\frac{cp}{W}\cos^2\theta. \tag{5}$$

Fig. 1a shows the normalized probability distribution of Eq. (3) as a function of $E$ and $\theta$, with a $^{60}$Co $\beta$ spectrum of endpoint energy $E_0 = 318$ keV used for d$P$/d$E$, for isotropic emission d$P$/dcos$\theta$ = 1. Fig. 1b shows separately the second term dcos$\theta$/d$t$ of Eq. (3), which plays a role only for high values of $E$ and low values of $\theta$.

However, we want the ToF spectra as a function of flight time $t$, and not as a function of the two independent variables $E$ and $\theta$. To this end, we resolve Eq. (1) for the variable $E$, which we write as a function of $\theta$ for a set of constant values of flight time $t_i$,

$$E_i = m_0 c^2 \left( \frac{c t_i \cos\theta}{\sqrt{c^2 t_i^2 \cos^2\theta - z_0^2}} - 1 \right). \tag{6}$$

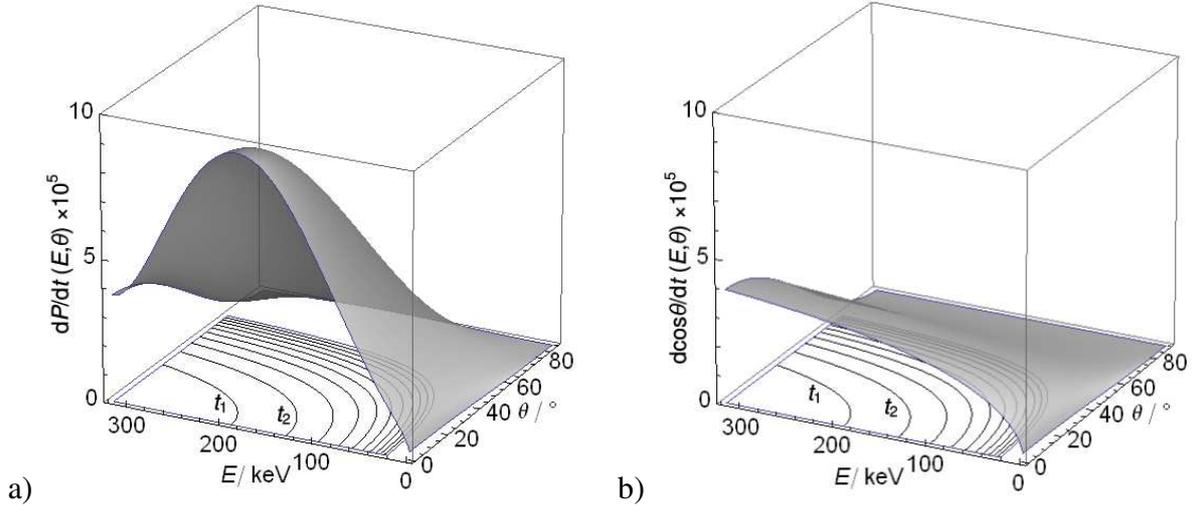

**Fig. 1.** a) The normalized ToF probability distribution (per degree and per keV) as a function of $E$ and $\theta$, calculated from Eq. (3) for the $\beta$ spectrum of $^{60}$Co decay. Curves of constant flight times are shown in the $E$-$\theta$ plane. The desired ToF spectrum then is obtained as a line integral of the probability distribution along these isochronous curves, see text. b) The second term of Eq. (3) is displayed separately, for isotropic electron emission d$P$/dcos$\theta$ = 1.

In the basic $E$-$\theta$ plane of Fig. 1, a family of such curves $E_i(\theta)$ is displayed. The $t_i$ therein increase, in steps of 10 ns, from $t_1 = 45$ ns up to $t_{10} = 145$ ns. This presentation of the curves $E_i(\theta)$ in Fig. 1 serves two purposes: First, each point of the desired ToF spectrum d$n$/d$t$ ($t$), is the line integral of the function d$P$/d$t$ ($E$,$\theta$) of Fig. 1a over one such isochronous lines $t$ = const in the $E$-$\theta$ plane. Each point in the ToF spectrum takes about 10 ms of CPU time on a laptop. Second, these isochronous curves verify explicitly what was known from the beginning. Namely, that for electron guidance in a magnetic field, a measured flight time value does not yield the energy $E$ of the electron (for our topic A) without knowledge of its emission angle $\theta$, nor (for our topic B) its emission angle $\theta$ without knowledge of its energy $E$.

The normalized ToF spectrum for $^{60}$Co $\beta$-decay resulting from these line integrals is shown in Fig. 2. The vertical dashed lines indicate the positions of a set of energies $E_i$ on the flight time $t$ axis, for electrons emitted under $\theta = 0$. For non-zero angles $\theta$, these energies move to larger values of $t$.

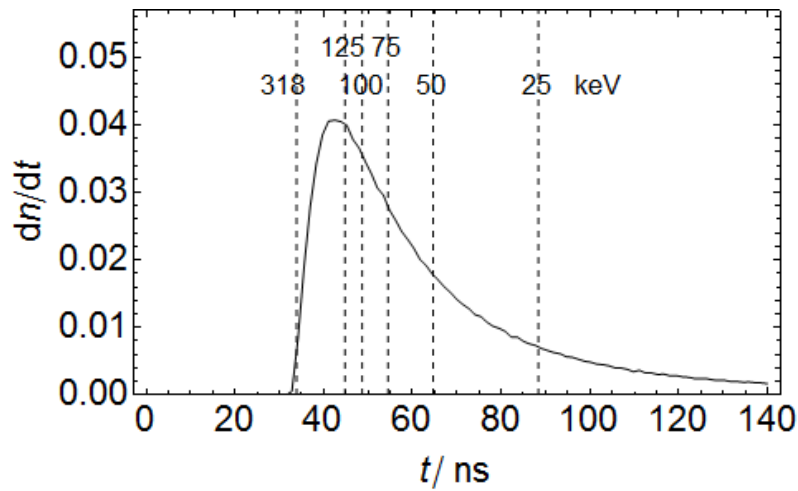

**Fig. 2.** Calculated normalized ToF spectrum for $^{60}$Co $\beta$-decay, in units ns$^{-1}$. The vertical dashed lines show several corresponding electron energies, increasing in steps of 25 keV, with an extra line for the endpoint energy $E_0 = 318$ keV.

## 3. Detector calibration with electron ToF in non-uniform magnetic fields

Various methods for energy calibration (A) are in use for electrons in the energy range of neutron decay, whose endpoint energy is 783 keV.

i) The most popular method is "in-situ" calibration with monochromatic electrons of known energies, as obtained from a set of conversion electron emitters like $^{109}$Cd, $^{139}$Ce, $^{113}$Sn, $^{137}$Cs, or $^{207}$Bi. These sources are installed within the spectrometer and are moved into the line of sight of the detectors in regular time intervals during data taking.

ii) This method can be accompanied by "off line" calibration with an uncalibrated broadband electron source and a magnetic momentum filter.

iii) Another off-line method uses Compton scattering of $\gamma$ radiation, creating electrons of known energy within the bulk of the detector [19].

Ideally, energy calibration should be independent of the quality of the electron source, which favours calibration with filtered electrons, method ii). On the other hand, this calibration should be done in situ, as in i), in order to monitor drifts or deterioration of detector parameters in the course of a running experiment. While Compton scattering, method iii), has its merits in the characterization of the bulk properties of detector material, it is insensitive to effects involving the surfaces of the detectors, like dead layers in silicon detectors or the development of surface cracks in plastic scintillators.

At present, a large new neutron decay spectrometer named PERC is being constructed by a Heidelberg-Grenoble-Munich-Vienna collaboration [20], to be installed at the FRM II reactor neutron source of the Technical University of Munich. A main feature of this instrument is the great length of its magnetic guiding field region. It consists of an 8 m long uniform region of low magnetic field (up to 2 T), and a short region of high field (up to 6 T). With this field configuration, electron ToF can become a useful tool for detector characterization, namely, for energy calibration of electron detectors (A), and for electron backscattering studies (B). In the PERKEO neutron decay asymmetry experiments [21-23], electron ToF cuts serve only to find the initial direction of electron emission in the case of backscattering events that give signals in both detectors.

The preceding section had shown that ToF measurements of electrons, guided to their detector by a *uniform* magnetic field, give no useful information on their energy. A solution to this problem is to make the electron's flight time independent of emission angle $\theta$. This is achieved by the "magnetic field parallelizer" method, also used in (the first half of) the KATRIN retardation spectrometer. This "magnetic bottle ToF method", as it is called by photoelectron spectroscopists (a better name is "inverse magnetic mirror ToF method"), had been proposed some decades ago [24,25] in the context of photoelectron spectroscopy. It turns out that the PERC instrument is well suited for this type of measurement, with a magnetic field profile similar to that shown in Fig. 3. In the design of PERC [26], special attention was given to the adiabatic condition in magnetic transport, which the electrons from neutron decay must meet, and which is also mandatory for the methods proposed in the present article.

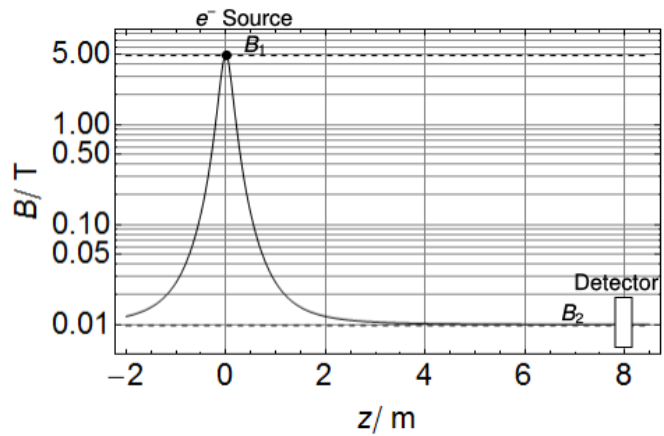

**Fig. 3.** The magnetic field profile $B(z)$ as used in the calculation of the electron ToF spectra with the inverse magnetic-mirror method. The calibration source is located in the peak field $B_1$, the detector is installed in the low-field section with field $B_2$ at 8 m distance from the source.

A broad-band electron source is placed at $z_1 = 0$, where the magnetic field $B(z)$ has its peak value $B_1 = 5$ T. The electron detector to be calibrated is installed at position $z_2 = 8$ m, where we let the field drop to a value $B_2 = 10$ mT. The extension of the calculation of the preceding section to the non-uniform field case is straightforward. By the inverse magnetic mirror effect, an electron, emitted into the direction of the detector, will on its way increase its longitudinal velocity component $v_z$, at the expense of its transverse components $v_x$ and $v_y$, its amplitude $v$ being a constant of the motion. Thereby, the angle $\theta = \arccos(v_z/v)$ between the velocity vector and the $z$-axis will continuously diminish on the electrons flight along $B(z)$. Under the adiabatic condition, this angle depends on the local field value $B(z)$ and the starting angle $\theta_0$ at field $B_1$, with

$$\sin\theta(z) = \sqrt{B(z)/B_1}\,\sin\theta(0)\,. \tag{7}$$

For the case of electron emission under right angles $\theta(0) = 90°$, the angle $\theta(z)$ (relative to the z-axis) rapidly diminishes along z, down to $\theta = 2.5°$ in the 500 times weaker final field $B_2 = 10$ mT, where $\cos\theta = 0.9990$.

In contrast, under adiabatic transport, the initial radius of electron gyration about $B(z)$, $r(0) = r_0 \sin\theta(0)$, increases as

$$r(z) = \sqrt{B_1/B(z)}\,r(0)\,, \tag{8}$$

up to a factor of $\sqrt{500} \approx 22$. For electrons of 1 MeV energy, the largest radius of gyration is $r_0 = 21$ mm at the detector position.

Fig. 4 shows $z_{\text{eff}}$ as a function of $\theta$ (full line), calculated for the field profile of Fig. 3 (from now on, we write simply $\theta$ for $\theta(z=0)$). The strong dispersion of path lengths met in the uniform field case (dashed line, diverging for $\theta \to 90°$) has almost completely disappeared.

With a non-uniform guiding field, the flight times in Eq. (1) must be replaced by the integral along axis z

$$t = \int_{z_1}^{z_2} \frac{dz}{v(z)} = \frac{1}{v}\int_{z_1}^{z_2} \frac{dz}{\sqrt{1-[B(z)/B_1]\sin^2\theta(0)}} \equiv \frac{z_{\text{eff}}}{v}\,, \tag{9}$$

in our case with $z_1 = 0$ and $z_2 = 8$ m. The calculation of the ToF spectra then proceeds in the same way as in Sect. 2 for a uniform field, but with the length $z_0/\cos\theta$ of the electron trajectory, with $z_0/\cos\theta$ (here $z_0 = z_2 - z_1$) replaced everywhere by $z_{\text{eff}} = v\,t$ from Eq. (9). Under the adiabatic condition, there is no need to bother with electron gyration in the x-y plane.

Fig. 5 shows the probability distribution for this non-uniform field case, which replaces Fig. 1a for the uniform field case. This distribution is less dependent on emission angle $\theta$, and the d$\cos\theta$/d$t$ term on the right hand side of Eq. (3) (not shown) has become completely negligible. Most important, however, is that the isochronous lines in the $E$-$\theta$ plane (along which the distribution must be integrated) have become nearly straight lines.

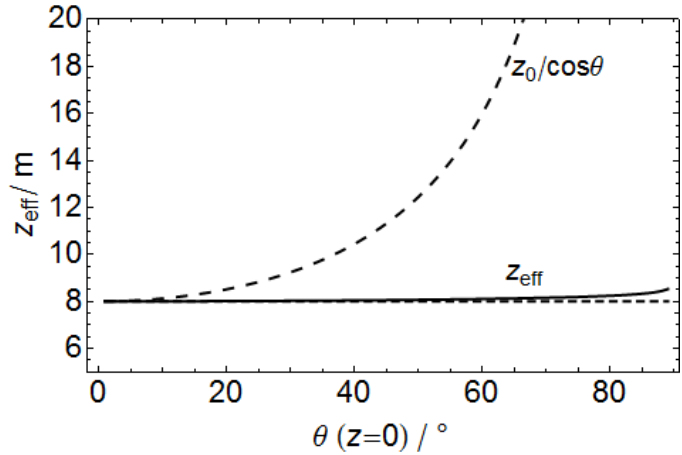

**Fig. 4.** For the rapidly decaying guiding field of Fig. 3, all electron path lengths $z_{eff}(z_2)$, shown as a function of initial emission angle $\theta$, become nearly the same, even for electron emission near $\theta = 90°$ (full line). In contrast, with a uniform guiding field, the lengths of the electron trajectories $z_0/\cos\theta$ diverge for $\theta = 90°$ (wide-dashed curve). Without a guiding field, the length of the flight paths has a constant value of 8 m (narrow-dashed line).

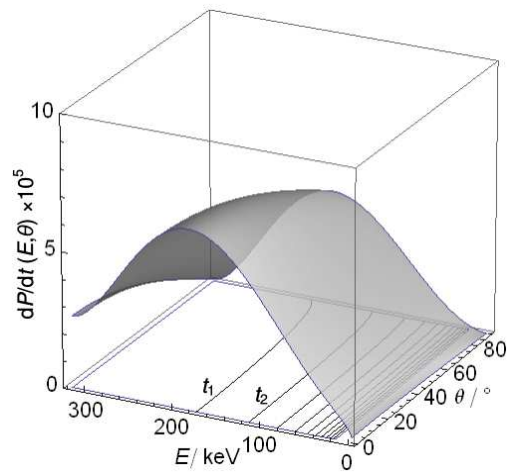

**Fig. 5.** Same ToF probability as in Fig. 1a, but for the non-uniform field $B(z)$ of Fig. 3. The distribution has smoothened, and the curves of constant ToF in the $E$-$\theta$ plane have become nearly straight lines.

The ToF spectrum of $^{60}$Co, calculated from this distribution is shown in Fig. 6. The spectrum has narrowed in comparison to the corresponding spectrum for a uniform guiding field, Fig. 2. In order to find out how well one can trust the electron energies derived from such a ToF spectrum, the calculation was repeated for a set of monoenergetic electrons, of energies $E = 50, 100, 300, 500$ keV, and 1 MeV. A finite intrinsic time resolution of the detection chain was taken into account. The presently operating PERKEO neutron decay spectrometer [27] uses plastic scintillators of type Bicron BC 404 for electron detection, with 2.2 ns pulse width. For such detectors, time resolutions of a few hundred picoseconds can be reached, and similar time resolutions are possible with multichannel plates, while lithium drifted silicon (SiLi) detectors have typical time resolutions of several nanoseconds [28].

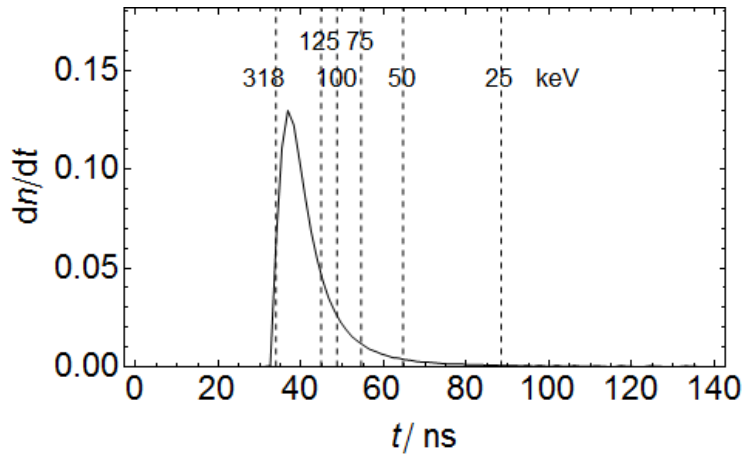

**Fig. 6.** The $^{60}$Co ToF spectrum for the non-uniform guiding field of Fig. 3. The vertical dashed lines have the same meaning as in Fig. 2.

In the calculation, a detector time resolution of ½ ns was assumed for electron energies greater than 500 keV. For the lower energies, this initial resolution was lowered, conservatively to the values listed in the second column of Table 1. Fig. 7 shows some calculated ToF spectra, whose characteristic parameters are listed in this table. Columns 3 to 5 of the table give, respectively, the flight times $t$ (at the peaks of the spectra), their widths $\Delta t$ (full widths at half maximum, FWHM), and the corresponding widths $\Delta E$ in energy. The widths $\Delta t$ in the 4$^{th}$ column are not much broader than the initial detector resolution in the second column. Below 500 keV, the relative energy resolution, column 6, is rather constant at $\Delta E/E \approx 13\%$. Shifts of the spectra (not shown in the table) are below 3%. The asymmetries shown in the column 7 are defined as the difference of half widths above and below energy $E$ from column 1, divided by the full widths $\Delta E$ from column 5. The spectra become very asymmetric only at energies $E$ of order 1 MeV.

| 1. | 2. | ToF: | | | | | Scintillator: |
|---|---|---|---|---|---|---|---|
| Electron $E$/keV | Intrinsic $t$-resol./ns | 3. $t$/ns | 4. $\Delta t$/ns | 5. $\Delta E$/keV | 6. $\Delta E/E$ | 7. Asym. | 8. $\Delta E/E$ |
| 50 | 3 | 65 | 3.5 | 6 | 12% | 8% | 66% |
| 100 | 2 | 49 | 2.2 | 12 | 12% | 8% | 47% |
| 300 | 1 | 34 | 1.2 | 42 | 14% | 10% | 27% |
| 500 | 0.5 | 31 | 0.7 | 70 | 14% | 15% | 21% |
| 1000 | 0.5 | 28 | 0.7 | 300 | 30% | 20% | 15% |

**Table 1.** Data derived from flight time calculations for monoenergetic electrons, Fig. 7.

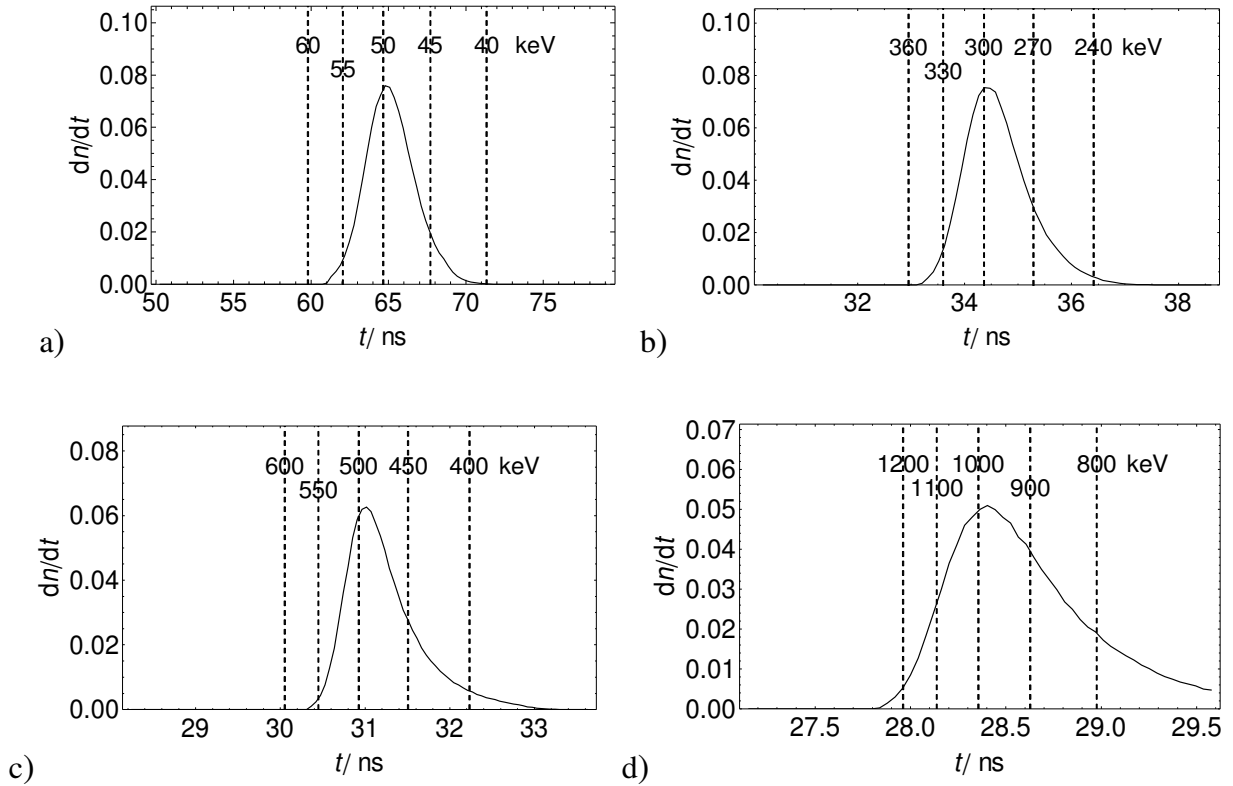

**Fig. 7.** ToF signals, calculated for the magnetic guiding field of Fig. 3, for monoenergetic electrons of energies $E = 50, 300, 500$ keV, and 1 MeV, for the intrinsic detector time resolutions listed in column 2 of Table 1. The vertical dashed lines indicate the positions of energy $E$, $E \pm 10\%$, and $\pm 20\%$ on the $t$ axis, from Eq. (6) for $\theta = 0°$.

For comparison, column 8 of Table 1 gives the relative widths for the calibration of plastic scintillators with conversion electrons, method i). Its values $\Delta E / E = 2\sqrt{2\ln 2}/\sqrt{N}$ (FWHM) are due to photoelectron statistics, where a realistic value of $N/E = 250$ photoelectrons per MeV electron energy is assumed. At and below 500 keV, the line widths from ToF in column 6 are considerably narrower than the widths in column 8 from conventional direct energy measurements.

As the line shapes of the ToF spectra for monoenergetic electrons in Fig. 7 are well known, the corresponding energies can realistically be determined to about 10% of the absolute line widths, column 5 in Table 1. For the lower part of the neutron decay spectrum, this translates into one-sigma calibration errors of one or a few keV. This would be a large improvement, compared to the present situation. The main advantage of the ToF method, however, remains that it provides us with an independent method of energy calibration that requires no a priori knowledge of the source spectrum.

Electron backscattering on the detector is negligible with this calibration method, because the magnetic mirror effect returns nearly all backscattered electrons back onto the detector, such that almost no energy is lost. The effect of backscattering can in any case be calculated using the results on electron backscattering (B), which are presented in the following section.

## 4. A complete ToF experiment of electron backscattering coefficients

Electron flight times can be used to measure the differential backscattering coefficients $\eta(\theta,\theta',E,E',Z)$ (B) for all angles $\theta$ and $\theta'$ and energies $E$ and $E'$ of the incoming and outgoing electrons, in one single measurement, for a given detector material of mean atomic number $Z$. The instrument setup, shown in Fig. 8, is similar to that of earlier PERKEO experiments [21,27], with two electron detectors connected here by a uniform guiding field. With PERKEO, electron ToF cuts had been used before to identify the spectrum of unrecognized backscattering events [29].

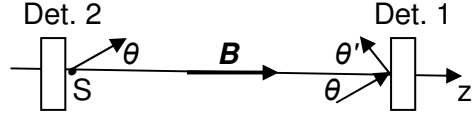

**Fig. 8.** Setup for the in-situ measurement of electron backscattering coefficients by ToF, with two electron detectors in a uniform field. The electron source S is placed near (or inside) detector 2, see Fig. 11.

A broadband electron source S is installed near detector 2. At time $t_0$, measured by one of the methods discussed in the next section, the source emits an electron of energy $E$ under polar angle $\theta$. The electron arrives at detector 1 at time $t_1$, with unchanged values of $E$ and $\theta$, is backscattered under angle $\theta'$ with energy $E'$, and reaches detector 2 at time $t_2$. Only events with the time sequence $t_0 < t_1 < t_2$ are registered. The backscattering coefficient then is the ratio $\eta = n_{12}/n_2$, where $n_{12}$ is the rate of coincidences between detectors 1 and 2, and $n_1$ the number of counts in detector 1.

To recall, in the preceding section on detector calibration (A) the electron flight times were nearly independent of angle $\theta$, and therefore energy $E$ could be directly derived from the measured flight time $t$ and be compared with the energy measured in the electron detector. In contrast, for backscattering studies (B), electron energies $E$ and $E'$ and flight times $t = t_1 - t_0$ and $t' = t_2 - t_1$, measured in the setup of Fig. 8, uniquely determine the angles $\theta$ and $\theta'$.

The initial total energy $E$ is measured as the sum of all energies deposited in detectors 1 and 2 (the same as in PERKEO). The energy of the backscattered electron is $E' = E - E_1$, where $E_1$ is the energy deposited and separately measured in detector 1. Methods to measure the start time $t_0$ will be discussed in the following section, together with some other questions.

With a long uniform guiding field, for instance the field in the "low-field region" of PERC (which can actually reach a value $B = 2$ T), all angles $\theta$ and $\theta'$ from 0 up to 90° become accessible. The highest accuracy is obtained for $\theta \approx 90°$ near glancing incidence, and the same for $\theta'$. Such glancing angles are difficult to handle both in conventional backscattering experiments and in computer simulations.

If only a shorter solenoid is available, electron flight times, and with it measurement accuracy, can be increased if the detectors are installed symmetrically in the fringe fields near the ends of the solenoid, where the field amplitudes $B_1 = B_2$ are lower than the solenoid's peak field $B_0$. The electrons arriving at detector 1 are required to surmount the magnetic barrier between the two detectors, and therefore their angular distribution is limited to angles below the critical angle of magnetic reflection,

$$\theta < \theta_{crit} = \arcsin(B_1 / B_0) \;, \tag{10}$$

see Fig. 9 for the case of a solenoid of 1 m length and 20 cm diameter.

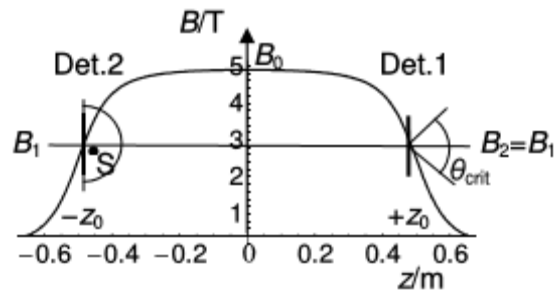

**Fig. 9.** If a short solenoid is used for the measurement of backscattering coefficients, electron flight times can be considerably prolonged (Fig. 10) by installing source S and detectors in the fringe fields of the solenoid.

For electron emission angles slightly below this critical angle, the electron flight times diverge, as shown in Fig. 10. If only events with flight times $t$ longer than a time $t_a$ are accumulated, up to some upper time limit $t_b$, or $t_a < t < t_b$, then only angles within the narrow band $\theta_a < \theta < \theta_b$ of corresponding angles are registered, see Fig. 10. By varying the positions of the detectors in Fig. 9 along $z$, this narrow band can be shifted to other values on the $\theta$ axis of Fig. 10. This change of detector positions can be avoided if (like in PERC) the peak field $B_0$ in Eq. (10) can be varied independently, without changing $B_1$ or $B_2$. The outgoing angles then are limited to values $0 \leq \theta' \leq \theta_{crit}$. Even this last limitation can be overcome to make all angles $0 \leq \theta, \theta' \leq 90°$ accessible, if asymmetric positions of the detectors are chosen, but I do not go into these details. These backscattering measurements can be done "in situ" during measurements with PERC, in this way testing the detectors in their actual state of quality.

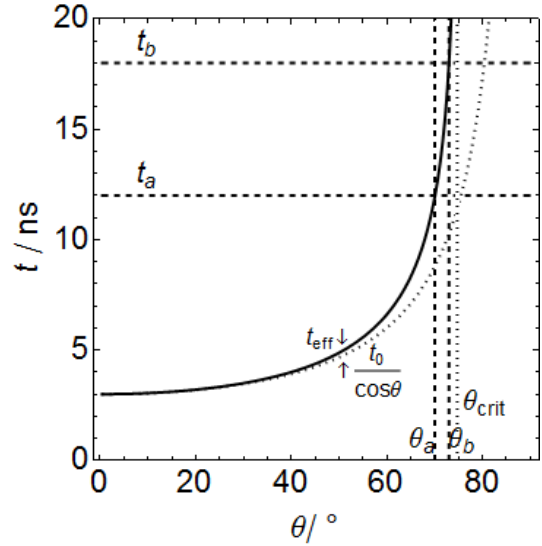

**Fig. 10.** Electron flight times vs. emission angle, for the setup of Fig. 9, for $E = 300$ keV. The flight times in $B(z)$ (full curve) are longer than the flight times in a uniform field (dotted curve). If long values of flight times $t_a < t < t_b$ are selected (dashed horizontal lines), this separates out electrons in a narrow band of emission angles $\theta_a < \theta < \theta_b$ (dashed vertical lines) near the critical angle $\theta_{\mathrm{crit}}$ (dotted vertical line), rather independent of energy.

## 5. Some technical issues

This section discusses some technical details, first on the magnetic fields. When PERC is installed on a neutron beam, its low-field region is upstream and may not be easily accessible. In this case, for detector calibration with method A, a separate low-field region can be installed downstream instead, consisting of a normal-conducting long solenoid with $B = 10$ mT. Electron energy spectroscopy by ToF can also be done without PERC, replaced altogether by a superconducting ring coil of 15 cm radius, plus the long conventional low-field solenoid. Actually, the field profile of Fig. 3 was calculated for such a set of current sources.

For the electron backscattering studies, method B, an ordinary superconducting solenoid of rather short length can be used instead of PERC, as was discussed in the preceding section. Note that the ToF spectra do not depend on the absolute value of the magnetic guiding field, as long as the adiabatic condition is met, and as long as the gyration radii of the electrons fit the size of the detector.

Next issues are the electron source, and the generation of the ToF start pulses. In both method A and B, energy is measured separately, and therefore the exact shape of the electron spectrum needs not be known. This permits using the emitted electrons themselves to produce the start signal, by placing the electron source inside detector 1, as indicated in Fig. 11a, which is easily realized for a plastic scintillator. In this case, any isotope that emits electrons in the desired energy range can be used, for instance the pure $\beta$ emitter $^{90}$Sr with a half life of 29 years and an endpoint energy of 2.3 MeV.

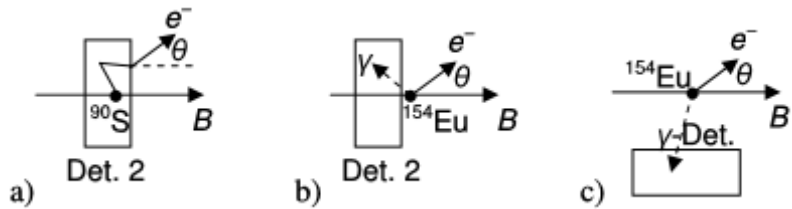

**Fig. 11.** Alternative ways of producing start signals for the ToF measurements.
a) By the electron itself, with the source inside a plastic scintillator as detector 2.
b) By a prompt $\gamma$ ray registered in detector 2. c) By a prompt $\gamma$ ray registered in a separate $\gamma$ detector.

A second method of generating a start pulse uses a source that emits both electrons and gamma rays in a prompt cascade, which is placed in front of detector 2, Fig. 11b. A possible isotope is $^{154}$Eu with a half life of 9 years. This isotope has many $\beta$ transitions, up to 1.8 MeV endpoint energy, followed in 89% of the decays by a $\gamma$ transition of energy 123 keV, from a level with 1.2 ns half life. A 10 mm thick plastic scintillator has 60% $\gamma$ detection efficiency at this energy. The two configurations shown in Figs. 11a and 11b are preferred for electron backscattering (B), where a second detector is needed anyway, installed near an end of the solenoid, where a scintillator plus photomultiplier are easily accommodated, as was shown in [17].

With ToF used for detector calibration (A), no second electron detector is needed, and the $\gamma$ detector can be installed off axis, as shown in Fig. 11c. In the high field region where the source is installed, a silicon $\gamma$ detector is preferable.

Concerning energy measurements in a backscattering experiment (B), the energy $E_1$ deposited in detector 1 must lie above the detector threshold. For a plastic scintillation detector, the threshold is typically 20 keV, which corresponds to five detected photoelectrons, for an energy sensitivity of 250 photoelectrons per MeV. This requirement may limit the detection of backscattering under glancing incidence to higher electron energies. Furthermore, measurement of total energy $E$ in both detectors, and separately of energy deposition $E_1$ in detector 1, requires precise electronic timing, in particular to cope with multiple backscattering events (which are of order $\eta^2$, typically several percent). Should these timing requirements be too ambitious for a first run, one can first drop the measurement of the second flight time $t'$ and determine only $\eta(\theta,E,E',Z)$, averaged over exit angles $\theta'$.

Dead time and accidental coincidences will not be a problem. Even after multiple backscattering, gate times for electron detection are shorter than one microsecond, and the permitted event rates are sufficient even if solid angle for $\gamma$ detection is well below $4\pi$.

## Conclusions

In the past, time-of-flight methods played no role in $\beta$-decay studies. The article shows that ToF can be a useful tool for in-situ characterisation of electron detectors in modern neutron decay spectrometers, and possibly for nuclear $\beta$-decay in general. In particular, electron ToF permits detector calibration with keV precision even for electron energies well below 100 keV where conventional calibration methods fail. Furthermore, backscattering on electron detectors can be studied, energy and angle resolved, down to low energies and to glancing incidence of the electrons.

## Acknowledgements

This work was supported by the Priority Programme SPP 1491 of Deutsche Forschungsgemeinschaft.

## References

[1] Barry R Holstein, Precision frontier in semileptonic weak interactions: theory,
    J. of Phys. G 41 (2014) 110301.
[2] Vincenzo Cirigliano, Michael J. Ramsey-Musolf, Low energy probes of physics beyond the standard model, Progr. Part. Nucl. Phys. 71 (2013) 2.
[3] Nathal Severijns, Oscar Naviliat-Cuncic, Symmetry tests in nuclear beta decay,
    Ann. Rev. Nucl. Part. Sci. 61 (2011) 23
[4] D. Dubbers, M.G. Schmidt, The neutron and its role in cosmology and particle physics,
    Rev. Mod. Phys. 83 (2011) 1111.
[5] H. Abele, Hartmut, The neutron. Its properties and basic interactions,
    Progr. Part. Nucl. Phys. 60 (2008) 1.
[6] J.S. Nico, W.M. Snow, Experiments in Fundamental Neutron Physics,
    Ann. Rev. Nucl. Part. Sci. 55 (2005) 27.
[7] C. Patrignani, et al. (Particle Data Group), Chinese Physics C, 40 (2016) 100001.
[8] J.W. Martin, J. Yuan, M.J. Betancourt, B.W. Filippone, S.A. Hoedl, T.M. Ito, B. Plaster, A.R. Young, New measurements and quantitative analysis of electron backscattering in the energy range of neutron β-decay, Phys. Rev. C 73 (2006) 015501.
[9] F. Wauters, I. Kraev, D. Zákoucky, M. Beck, V.V. Golovko, V.Yu. Kozlov, T. Phalet, M. Tandecki, E. Traykov, S. Van Gorp, N. Severijns, A GEANT4 Monte-Carlo simulation code for precision b spectroscopy, Nucl. Instrum. Methods A 609 (2009) 156
[10] D. Dubbers, L. Raffelt, B. Märkisch, F. Friedl, and H. Abele,
    Nucl. Instr. Meth. A 763 (2014) 112.
[11] A.M. Rijs, E.H.G. Backus, C.A. de Lange, N.P.C. Westwood, M.H.M. Janssen,
    'Magnetic bottle' spectrometer as a versatile tool for laser photoelectron spectroscopy,
    J. Electron Spectroscopy and Related Phenomena 112 (2000) 151.
[12] J.F. Amsbaugh, et al., Focal-plane detector system for the KATRIN experiment,
    Nucl. Instrum. Methods A 778 (2015) 40.
[13] E.W. Otten, C. Weinheimer, Neutrino mass limit from tritium β decay,
    Rep. Prog. Phys. 71 (2008) 086201.
[14] O. Cheshnovsky, S. H. Yang, C.L. Pettiette, M. J. Craycraft, R. E. Smalley, Magnetic time-of-flight photoelectron spectrometer for mass-selected negative cluster ions, Rev. Sci.lnstrum. 56 (1987) 2130.

[15] V. Lollobrigida, G. Greco, D. Simeone, F. Offi, A. Verna, G. Stefani, Electron trajectory simulations of time-of-flight spectrometers for core level high-energy photoelectron spectroscopy at pulsed X-ray sources,
J. Electron Spectroscopy and Related Phenomena 205 (2015) 98.
[16] D. Dubbers, Magnetic guidance of charged particles, Phys. Lett. B 748 (2015) 310.
[17] Dirk Dubbers, Ulrich Schmidt, Generation of narrow peaks in spectroscopy of charged particles, Nucl. Instrum. Methods A 837 (2016).50
[18] Alexander Osipowicz, Bernhard Zipfel, Electron optical imaging properties of the KATRIN high field solenoid chain, Nucl. Instrum. Methods A 760 (2014) 68.
[19] C. Aberle, C. Buck, F.X. Hartmann, S. Schönert, S. Wagner, Light output of Double Chooz scintillators for low energy electrons, 2011 JINST 6 P11006.
[20] G Konrad, et al., Neutron decay with PERC: A progress report,
J. of Phys. Conf Ser. 340 (2012) 012048.
[21] D. Mund, B. Märkisch, M. Deissenroth, J. Krempel, M. Schumann, H. Abele, Determination of the weak axial vector coupling $\lambda = g_A/g_V$ from a measurement of the -asymmetry parameter $A$ in neutron beta decay, Phys. Rev. Lett. 110 (2013) 172502.
[22] M. Schumann, T. Soldner, M. Deissenroth, F. Glück, J. Krempel, M. Kreuz, B. Märkisch, D. Mund, A. Petoukhov, H. Abele, Measurement of the neutrino asymmetry parameter $B$ in neutron decay, Phys. Rev. Lett. 99 (2007) 191803.
[23] M. Schumann, M. Kreuz, M. Deissenroth, F. Glück, J. Krempel, B. Märkisch, D. Mund, A. Petoukhov, T. Soldner, H. Abele, Measurement of the proton asymmetry parameter in neutron beta decay, Phys. Rev. Lett. 100 (2008) 151801.
[24] G. Beamson, H.Q. Porter, D.W. Turner, The collimating and magnifying properties of a superconducting field photoelectron spectrometer, J. Phys. E 13 (1980) 64.
[25] P. Kruit, F.H. Read, Magnetic field paralleliser for 2π electron-spectrometer and electron-image magnifier, J. Phys. E 16 (1983) 313.
[26] D. Dubbers, H. Abele, S. Baeßler, B. Märkisch, M. Schumann, T. Soldner, O. Zimmer, A clean, bright, and versatile source of neutron decay products,
Nucl. Instrum. Methods A 596 (2008) 238.
[27] B. Märkisch, H. Abele, D. Dubbers, F. Friedl, A. Kaplan, H. Mest, M. Schumann, T. Soldner, D. Wilkin, The new neutron decay spectrometer PERKEO III,
Nucl. Instrum. Methods A 611 (2009) 216.
[28] G.F. Knoll, *Radiation Detection and Measurement*, 4[th] ed., Wiley, Hoboken, N.J., 2010.
[29] M. Schumann, H. Abele, Unrecognized backscattering in low energy beta spectroscopy, Nucl. Instrum. Methods A 85 (2008) 88.